\documentclass{article}
\usepackage{graphicx, caption, amsmath}
\usepackage[margin=1in]{geometry}
\usepackage{setspace}
\usepackage{authblk}
\usepackage{xcolor}         
\usepackage{booktabs}       
\usepackage[hyphens]{url}   
\DeclareCaptionLabelFormat{nospace}{#1#2}

\newcommand{\Q}{\mathbf{Q}}
\newcommand{\Sd}{\text{$S^D$}}

\newcommand{\clr}{\text{CLR}}
\newcommand{\ilr}{\text{ILR}}
\newcommand{\alr}{\text{ALR}}

\newcommand{\tr}{^{\sf T}}

\begin{document}
\title{A Guideline for the Statistical Analysis of Compositional Data in Immunology}
\author[1]{\small Jinkyung Yoo}
\author[2]{\small Zequn Sun}
\author[3]{\small Michael Greenacre}
\author[4]{\small Qin Ma}
\author[4, *]{\small Dongjun Chung}
\author[1, *]{\small Young Min Kim}

\affil[1]{\footnotesize Department of Statistics, Kyungpook National University, South Korea}
\affil[2]{\footnotesize Department of Preventive Medicine - Biostatistics, Northwestern University, USA}
\affil[3]{\footnotesize Department of Economics and Business, Universitat Pompeu Fabra and Barcelona School of Management, Spain}
\affil[4]{\footnotesize Department of Biomedical Informatics, The Ohio State University, USA}
\affil[*]{\footnotesize To whom correspondence should be addressed (chung.911@osu.edu, kymmyself@knu.ac.kr).}

\vspace{2ex}

\date{}

\onehalfspacing
\normalsize
\maketitle
\abstract{
The study of immune cellular composition has been of great scientific interest in immunology because of the generation of multiple large-scale data. From the statistical point of view, such immune cellular data should be treated as compositional. In compositional data, each element is positive, and all the elements sum to a constant, which can be set to one in general. Standard statistical methods are not directly applicable for the analysis of compositional data because they do not appropriately handle correlations between the compositional elements. In this paper, we review statistical methods for compositional data analysis and illustrate them in the context of immunology. Specifically, we focus on regression analyses using log-ratio transformations and the generalized linear model with Dirichlet distribution, discuss their theoretical foundations, and illustrate their applications with immune cellular fraction data generated from colorectal cancer patients.
}

\section{Introduction} \label{sec1}

\begin{figure}
    \centering
    \includegraphics[width = .99\textwidth]{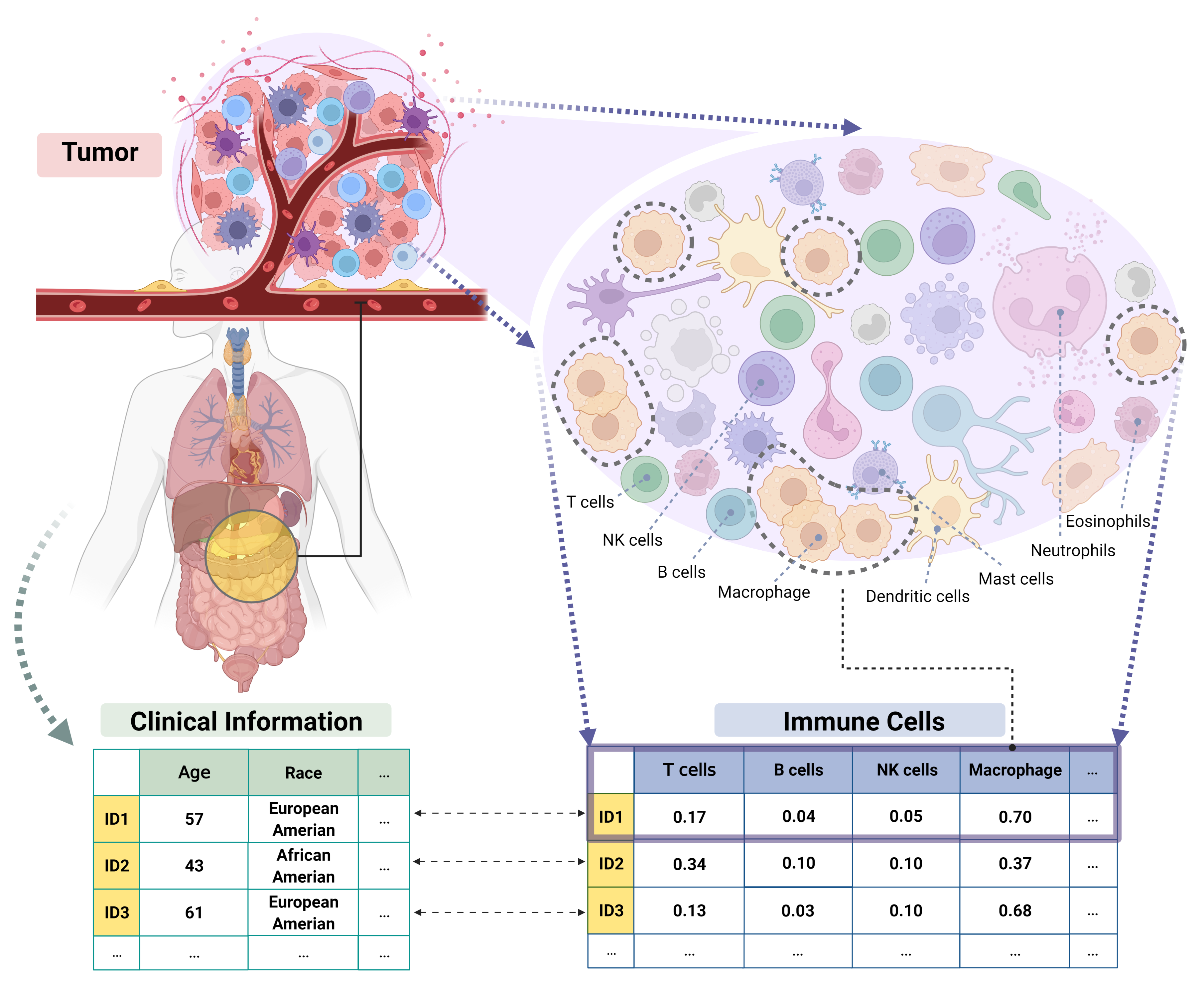}
    \caption{Immune Landscape of Cancer data and immune cell type composition within the tumor microenvironment of each cancer patient}
    \label{fig:immune_cancer}
\end{figure}

   The human immune system consists of various types of immune cells (e.g., T cells, B cells, natural killer (NK) cells,  dendritic cells,  among others). Upon viral infection, tissue transplantation, or disease occurrence, dynamic and extensive interaction among these immune cell types occurs in the human body. Hence, in the study of the human immune system, it is of great interest to understand composition, differentiation, and activities of various types of immune cells, and interactions among them. The composition of these immune cells is also associated with cancer progression, adverse events, and response to cancer immunotherapy, especially immune checkpoint blockades including Anti-PD1 and Anti-CTLA4. 
   
   Along with the interest in this association, there is a movement to gather immune cellular information and clinical information together. In the immunology field, multiple types of assays are used to interrogate such immune cellular composition, including flow cytometry and single cell RNA-seq. In addition, multiple computational algorithms have also been proposed to estimate immune cellular composition by deconvolving bulk gene expression data, where popular algorithms include CIBERSORT \cite{newman2019determining}. Recognizing the importance of understanding the immune cellular composition, and the emergence of these computational algorithms and relevant assays, have led to the generation of multiple large-scale immune cellular composition datasets. For example, the Immune Landscape of Cancer generated a large-scale immuno-genomic dataset from more than 10,000 patients with 33 different cancer types based on the Cancer Genome Atlas (TCGA) data \cite{thorsson_immune_2018}.
   Figure~\ref{fig:immune_cancer} illustrates the Immune Landscape of Cancer, which provides information about immune cell composition within the tumor microenvironment of each cancer patient, along with corresponding clinical information.
   This new type of data motivates the investigation of relevant statistical methods that can consider key characteristics of these datasets. Effective analysis of such datasets can potentially support development of diagnosis and treatment strategies for various diseases such as cancer and autoimmune diseases.
   
   From the statistical point of view, such immune cellular data can be considered as {\it compositional}, since it carries relative information only in the form of proportions of a total amount, summing to 1 in each sample. John Aitchison was a pioneer in the statistical formulation of compositional data analysis, and developed the relevant geometry, metrics, and a guideline for the application of various statistical methods in this context. The constituent cellular fractions are defined on the Aitchison simplex, not in Euclidean space. The Aitchison simplex, \Sd, is a sample space for compositional data, and is defined as
    \begin{align}
        \Sd = \left\{ (v_1, v_2, \dots, v_D) : v_d \geq 0, d=1,\dots,D, \sum_{d=1}^D v_d = 1 \right\},
        \label{simplex}
    \end{align}
    where $v_1,\dots,v_D$ are non-negative components of a {\it $D$-part composition}. The dimension of \Sd~is $D-1$ due to the constant sum constraint. Aitchison introduced  statistical methods based on log-ratios, which are still most popularly used to analyze compositional data (in this case the compositional parts in (\ref{simplex}) should be strictly positive) \cite{aitchison_1986}.
   The method is invariant to scaling of compositions, called {\it scale invariance}, which gives coherent results regardless of multiplication of a row (composition) of the initial data by an arbitrary positive constant.  The R environment includes several packages developed for the log-ratio approach, for example, \texttt{compositions} \cite{van2008compositions}, \texttt{robCompositions} \cite{templ2011robcompositions}, and \texttt{easyCODA} \cite{greenacre2018compositional}.
   
    Maier asserted that it is often not straightforward to interpret the results from data analyses using log-ratio transformations and, in addition, these methods can often violate modeling assumptions such as homoscedasticity \cite{dirichletR}. As an alternative approach, Dirichlet regression was proposed,
   originally suggested as a null model for compositional data by Campbell and Mosimann \cite{campbell1987multivariate}. Hijazi and Jernigan developed the maximum likelihood estimation methods for Dirichlet regression and also investigated the sampling distributions of the estimates \cite{hijazi2009modelling}. Camargo {\it et al.} introduced a new approach for estimating the Dirichlet model when each parameter has a linear structure on covariates and suggested a Bayesian model selection method \cite{camargo2012estimation}.
   
   These ongoing discussions are to determine optimal statistical strategies for compositional data analysis. Following the trend, in this paper, we aim to give a guideline for the statistical approaches of compositional data analysis in the context of immunology data: firstly, modeling using standard regression analysis with log-ratio transformations, and secondly, the approach using Dirichlet regression analysis.
   
   This paper is structured as follows. In Section 2 we introduce the immune cellular fractions data for colorectal cancer, and the two compositional regression approaches, log-ratio regression and Dirichlet regression, which have been applied to this dataset. Section 3 gives the results of these alternative modeling approaches. Section 4 summarizes the key findings of this paper and comments on the similarities and differences of the two approaches.
\section{Material and Methods}
\label{sec2}
\subsection{Colorectal cancer data}
\label{subsec2.1}
    In this paper, we focus on the analysis of the immune cellular fractions data of colorectal adenocarcinoma patients, generated from the Immune Landscape of Cancer project \cite{thorsson_immune_2018}. 
    Specifically, there are 254 colorectal adenocarcinoma patients, 58 (23\%) of which are African Americans (AA) and 196 (77\%) are European Americans (EA). These patients are almost equally divided between females and males, 126 and 128, respectively. Motivated by previous studies showing significant racial disparity in clinical outcomes \cite{king_thomas_racial_2019, curran_differential_2021}, we focus here on investigating associations of immune cell compositions with race. In the Immune Landscape of Cancer, the immune cellular fractions were estimated by deconvolving the gene expression data of the TCGA PanCancer study using the CIBERSORT algorithm \cite{newman_robust_2015}. Thorsson {\it et al.} provided three different aggregations of immune cell types, of which we use {\it Aggregate 2} \cite{thorsson_immune_2018}. Specifically, it consists of nine immune cell types: CD8\textsuperscript{+} T cells (labeled as {\tt T.cells.CD8} or {\tt T.CD8}), CD4\textsuperscript{+} T cells ({\tt T.cells.CD4} or {\tt T.CD4}), B cells ({\tt B.cells} or {\tt B}), NK cells ({\tt NK.cells} or {\tt NK}), macrophage ({\tt Macrophage} or {\tt Macro}), dendritic cells ({\tt Dendritic.cells} or {\tt Dendr}), mast cells ({\tt Mast.cells} or {\tt Mast}), neutrophils ({\tt Neutrophils} or {\tt Neutr}), and eosinophils ({\tt Eosinophils} or {\tt Eosin}).

\subsection{Log-ratio approaches}
\label{subsec:logratio}
   In this section, we describe how to apply the log-ratio regression model to the colorectal cancer data. Since this approach involves computing logarithms of ratios, zero values need to be replaced, which can be done in several ways \cite{lubbe2021comparison}.  
   In this study we used the k-nearest neighbours (KNN) approach of Hron {\it et al.} \cite{hron2010imputation}. 
   
   In the context of colorectal cancer data, we are mainly interested in racial differences in immune cellular compositions. 
   To investigate this relationship using log-ratio regression models, the  immune cellular compositions composed of the $D=9$ immune cells are used as multivariate responses and race and/or age as explanatory variables. 
   
    The key step in the log-ratio approach is to choose a set of log-ratio transformations, which convert all the compositions on the Aitchison simplex to multivariate vectors on interval scales in a regular Euclidean space.
    We can then proceed to apply standard statistical analyses to the log-ratio transformed data, with some care taken in the way the results are interpreted.
    The widely used sets of log-ratio transformations are the additive log-ratios (ALRs) \cite{aitchison82}, the centered log-ratios (CLRs) \cite{aitchison82}, and isometric log-ratio (ILRs) \cite{egozcue03isometric}. 
    All of these transformations can be written as $\log({\bf v}) \cdot {\bf Q}\tr$, where $\log({\bf v})$ is the row vector of compositional values transformed by the natural logarithm, and $\bf Q$ is a matrix with each row summing to zero --- see, for example, Greenacre (2022) \cite{greenacre2022compositional}.
    
    The ALR transformations are the easiest to understand and interpret, since they can be simply calculated by taking $D-1$ pairwise logratios with respect to a fixed denominator part, taken here as the last part:
    \begin{equation}\label{eq:alr}
        \textrm{ALR}(\mathbf{v}) = 
        \left( \log \frac{v_1}{v_D}, \dots, \log \frac{v_{D-1}}{v_D} \right) = \log({\bf v})\cdot {\bf Q}_\textrm{ALR}\tr
    \end{equation}
    where the $(D-1)\times D$ matrix $\bf Q_\textrm{ALR}$ is the $(D-1)\times (D-1)$ identity matrix with an additional $D$-th column of all $-1$ values. 
    Although technically the \alr~transformations 
    do not preserve exact isometricity, the denominator can be chosen to give a transformation close to being isometric \cite{greenacre2021etal}. 
    An isometric transformation is one that engenders the exact logratio geometry of all pairwise logratios (see, for example, Greenacre's work in 2019 \cite{greenacre2019variable}).
    The \alr~transformation has a intuitive interpretation due to its simple definition, which is advantageous for practical applications.

    The CLR transformations are defined as 
    \begin{equation}
    \label{eqn:clr}
        \textrm{CLR}(\mathbf{v}) = 
        \left( \log \frac{v_1}{g(\mathbf{v})}, \dots, \log \frac{v_D}{g(\mathbf{v})} \right)= \log({\bf v})\cdot {\bf Q}_\textrm{CLR}\tr \textrm{,\ where\ }  g(\mathbf{v}) \textrm{\ is\ the\ geometric\ mean}\sqrt[D]{v_1 \cdot v_2 \cdots v_D}.
    \end{equation}
    The $D\times D$ matrix $\bf Q_\textrm{CLR}$ has all off-diagonal values equal to $-\frac{1}{D}$ and diagonal values equal to $1-\frac{1}{D}$.
    The $D$ CLR transformations are symmetric with respect to the compositional components and they are isometric. The fact that the set of CLRs has a singular covariance matrix \cite{egozcue03isometric} is of no consequence to our present purpose of performing log-ratio regression. 

    Finally, the \ilr~transformations are the most complicated choice, both to define and interpret. It is also defined as a linear transformation
    \begin{equation}\label{eq:ilr}
        \textrm{ILR}(\mathbf{v}) = \log(\mathbf{v}) \cdot \Q_\textrm{ILR}\tr,
    \end{equation}
    where the $(D-1) \times D$ matrix $\Q_\textrm{ILR}$ has rows that contrast groups of compositional parts as ratios of their geometric means, along with an additional scalar multiplier. 
    This set of transformations is used when the interpretability of the transformation is of less importance than its properties of having a nonsingular covariance matrix and being isometric.
    A special case of ILR transformations is the set of pivot log-ratios \cite{filzmoser2018applied}, where single components are in the numerator of each ratio and the geometric means of the remaining components in the particular ordered sequence are in the denominator. 
    The problem here would be to decide what that ordered sequence should be, since there are $D!$ orderings.
    Again it should be noted that for the present purpose of compositional regression, the final results will be equivalent no matter which log-ratio transformation is chosen (see below, when the result is expressed as a log-contrast). Hence, if simplicity and interpretability are regarded as important, then the ALR transformations will be preferred.

    When we use the \alr~transformation, we need to decide which denominator part to choose.
    Since this choice will not affect our results, the fixed denominator can be chosen on substantive grounds to make the interpretation of the results more meaningful, or it can be chosen to optimize some favourable property. 
    For example, Greenacre {\it et al.} showed how to choose the denominator that gives the set of transformations closest to being isometric \cite{greenacre2021etal}.
    The {\it Procrustes correlation} \cite{gower_procrustes_2004,legendre2012numerical} can be used to measure the similarity of the geometry based on an \alr-transformation, with only $D-1$ variables, to the geometry based on all $\frac{1}{2}D(D-1)$ pairwise log-ratios \cite{greenacre2019variable}, sometimes called the Aitchison geometry. The same correlation can be used if an even smaller subset of pairwise log-ratios is selected and we want to measure how close to isometry they are, since it often happens that less log-ratios can be used to make the results more parsimonious \cite{greenacre2019variable, graeve2020selection}.
    
    As said above, once the data have been log-ratio transformed, then standard statistical analysis can be used. 
    Thus, multiple regression models on the log-ratio transformed data can be written as follows, for the $j$-th log-ratio ${\rm LR}_j$ of a set of transformations for 
    $j=1,2,\dots,D-1$ or $D$,  where LR can be ALR, ILR or CLR:
    \begin{equation}
        {\rm LR}_j({\mathbf v}) = \alpha_j + \beta_j x_1 + \gamma_j x_2 + {e_j}
        \label{eq:logreg0}
    \end{equation}
  where ${\alpha}_j$, $\beta_{j}$ and $\gamma_j$ are regression coefficients, $x_1, x_2$ are the two explanatory variables, race (dummy variable) and age in this case, and $e_j$ is a random error with mean $0$ and a constant variance. 
  
  Once the set of log-ratio regressions (Equation \eqref{eq:logreg0}) is performed for each log-ratio $j$, then the coefficients $\beta_j$ and $\gamma_j$ for the two predictors can be inversely transformed to give a \textit{log-contrast} on $\bf v$ of the form (for each predictor) \cite{van2013analyzing,greenacre2022compositional}.
  \begin{equation}
  \phi_1 \log(v_1) + \phi_2 \log(v_2) + \cdots + \phi_D \log(v_D), \quad \textrm{where\ } \sum_j \phi_j = 0.
  \label{eq:logcontrast}
  \end{equation}  
  This result shows how the compositional response, on a log-scale, is affected by a unit change in the predictor, as estimated by the set of log-ratio regressions.
  In the case of the dummy variable predictor for race, where EA = 0 and AA = 1, the coefficients show the ``effect" of AA compare to EA. 
  When Equation \eqref{eq:logcontrast} is exponentiated, the $\phi_j$'s are seen to be the estimated multiplicative effects, since the exponentiated log-contrast has this form: $v_1^{\phi_1} v_2^{\phi_2} \cdots v_D^{\phi_D}$.
  The coefficients $\phi_j$ of the log-contrast will be the same no matter which log-ratio transformation, \alr, \clr or \ilr, is used. 
  This further justifies using the \alr~ transformation, which not only produces the correct log-contrast result but also has an easier interpretation of the responses in the original log-ratio regressions (Equation \eqref{eq:logreg0}).
  When log-ratios are used as predictors, there is a similar result --- no matter which complete set of log-ratios is used, they all lead to the same log-contrast \cite{coenders2020interpretations}.
  
  When it comes to visualizing compositional data by a dimension-reduction method such as principal component analysis (PCA), then CLR-transformed data are used, since the PCA of the CLRs has been shown to be equivalent to the PCA of all pairwise log-ratios \cite{aitchison2002biplots}. 
  The PCA of CLR-transformed data has thus been called \textit{log-ratio analysis} (LRA) \cite{greenacre2010log}. Here there is an issue whether to weight the parts or not, which is usually decided when comparing the variance contribution of the parts to the total variance, since it often occurs -- as in this application -- that rare parts contribute excessively to the total variance.
  With weights $w_d, d=1,\ldots, D$, equal to the average compositional proportions, the total variance is then computed equivalently in two different ways as follows:
    \begin{align}
         \label{eq:TLRV}
         {\rm TotVar} &= \sum_{d} w_d \sigma^2_{d}  \\
               &= \sum \sum_{d<h}w_d w_h \sigma^2_{d,h} \nonumber
    \end{align}
    using either the variances $\sigma^2_{d}$ of the CLRs or the variances $\sigma^2_{d,h}$ of the pairwise logratios of the $d$-th and $h$-th parts \cite{greenacre2021compositional}. 


\subsection{Dirichlet approaches}
\label{subsec2.3}

\medskip
\leftline{\it The Dirichlet distribution}
\smallskip
   The Dirichlet distribution models the probability of a multinomial random variable, that is a random composition ${\bf v} = (v_1, v_2,\ldots,v_D)$ where $v_d \in [0,1]$ and $\sum_d v_d = 1$.
   The Dirichlet distribution has shape parameters  $\boldsymbol{\alpha} = (\alpha_1, \dots,\alpha_D)\tr$ for the $D$ respective components. The probability density function (pdf) is defined as
    \[
        \mathcal{D}(\mathbf{v}|\boldsymbol{\alpha}) = \frac{1}{B(\boldsymbol{\alpha})} \prod_{d=1}^D v_d^{\alpha_d -1}, \quad \text{where}\quad  {B}(\boldsymbol{\alpha}) = \frac{\prod_{d=1}^D \Gamma(\alpha_d)}{\Gamma(\sum_{d=1}^D \alpha_d)},   \]
   where ${B}(\cdot)$ and $\Gamma(\cdot)$ are the Beta and Gamma functions, respectively.
   The marginal distributions of the Dirichlet distribution are all Beta distributions.
   For each component $d$, the mean and variance of the marginal Beta distribution are $E[v_d] = \alpha_d/\alpha_{+}$ and $Var[v_d] = [\alpha_d(\alpha_{+} - \alpha_d)]/[\alpha_{+}^2(\alpha_{+}+1)$], where $\alpha_{+} = \sum_d \alpha_d$. $\alpha_{+}$ is a measure of dispersion (or precision) of the distribution, with high values of $\alpha_{+}$ indicating higher density around the expected values.
    
    \medskip
    \leftline{\it Estimation of Dirichlet regression coefficients}
    \smallskip
     \noindent The Dirichlet regression provides an alternative to log-ratio regression for the modeling of compositional data responses, and
    establishes the relationships between the parameters of the Dirichlet distribution and linear functions on the covariates. The regression model should be fitted for each $\alpha_d$.
    Suppose that there are independent random vector variables $V_1, \dots, V_n$, where $V_i = (V_{i1},\dots,V_{iD})$ for $i=1,\dots,n$ satisfying $\sum_{d=1}^D v_{id} = 1$ for observed $v_{id}$ for $V_{id}$. In addition, we assume that  given the covariate row vector $X_i = (x_{i1}, \dots, x_{ip})$ corresponding to $i$-th observation,  $V_i|X_i$ follows the Dirichlet distribution with shape parameters $(\alpha_{i1},\dots,\alpha_{iD})$, each of which satisfies for observed covariates $\mathbf{x}_i$,
    \begin{align*}
        \alpha_{id} = h_d(\beta_{d1}x_{i1} + \cdots + \beta_{dp}x_{ip}) = h_d(\boldsymbol{\beta}_d\tr \mathbf{x}_i)
    \end{align*}
    where $\boldsymbol{\beta}_d = \left\{ \beta_{dk} \right\}$, $k=1,\ldots,p$, is a vector of regression coefficients. Here, each function $h_d : R \rightarrow (0,\infty)$ is an three times differentiable injective function. The functions $h_d$ work in the analogous way that the link function in generalized linear models does. We also assume that $h_d = h$, especially designated by the log function \cite{gueorguieva2008dirichlet, melo2009some, dirichletR}. Therefore, the relation can be re-written for each element of $\alpha_d = \left\{ \alpha_{id} \right\}$ as
    \begin{align*}
        \log(\alpha_{id}) = \boldsymbol{\beta}_d\tr \mathbf{x}_i, \quad i=1,2,\dots, n.
    \end{align*}
    The unknown regression coefficients $\hat{\beta}_{dk}$ are estimated by using maxim likelihood method, through the derivatives of the log-likelihood function given by,
    \begin{equation*}
        L(\mathbf{v}|\tilde{\alpha}) = \sum_{i=1}^n \left\{ \log \Gamma\left(\sum_{d=1}^D \alpha_{id}\right) - \sum_{d=1}^D \log \Gamma(\alpha_{id}) + \sum_{d=1}^D (\alpha_{id} -1)\log v_{id}. \right\}
    \end{equation*}
   There is no closed form solution, hence it must be calculated numerically using a nonlinear optimization procedure. The invariance property of the maximum likelihood estimator (MLE)  leads to obtain the MLE $\hat{\alpha}_{id}$ of $\left\{ \alpha_{id} \right\}$ as
    \begin{align} \label{equ:mle_alpha}
        \hat{\alpha}_{id} = \exp(\hat{\boldsymbol{\beta}}_{d}\tr \mathbf{x}_i).
    \end{align}
    In this manuscript, we employed the Dirichlet regression for analysis of the colorectal cancer immunology data, where immune cellular composition is considered as conpositional outcomes. 
    
    \medskip
    \leftline{\it Dirichlet regression diagnostics}
    \smallskip 
    \noindent For model diagnostics of the Dirichlet regression, two types of residuals can be considered, namely standardized residuals and composite residuals. For the $d$-th component and the $i$-th individual, standardized residuals $r_{id}$ and composite residuals $C_i$ are defined as 
    \begin{align}
       \label{eq:SR} & r_{id} = \frac{v_{id} - E[V_{id}|\hat{\alpha}_{i \cdot}] }{ \sqrt{Var(V_{id}|\hat{\alpha}_{i \cdot})} }, \quad \mbox{and} \\
       \label{eq:CR} &  C_i = \sum_d r_{id}^2, 
    \end{align}
    for $d=1,2,\dots,D$, and $i=1,2,\dots,n$. Here, $V_{id}$ is a random variable for $v_{id}$ for $d=1,\dots,D$ and $i=1,\dots,n$, and $\hat{\alpha}_{i \cdot}$ is a $D\times 1$ vector as  $(\hat{\alpha}_{i1},\dots,\hat{\alpha}_{iD})$ defined in Equation \eqref{equ:mle_alpha}. 
    The composite residuals Equation \eqref{eq:CR} are obtained using equal weights to all standardized residuals Equation \eqref{eq:SR}. 
  
    For the Dirichlet regression model, Gueorguieva {\it et al.} investigated its diagnostic approach to identify influential observations utilizing score residuals, which can assess overall model fit through overdispersion \cite{gueorguieva2008dirichlet}. First of all, Cook suggested the {\it local measures of influence} to detect observations with high leverages \cite{cook1986assessment}. Specifically, the local measures are constructed by assigning minimal weight to an observation in the likelihood \cite{cook1986assessment}, whereas the original Cook's distance, which is used for the global measures of influence, deletes an observation completely. The measure is defined as 
    $$
        \rho_{id} = \frac{s_{id}}{G'(\hat{\alpha}_{id}) - G'(\sum_i \hat{\alpha}_{id})}, \quad d=1,2,\dots,D, \quad i=1,2,\dots,n,
    $$
    where $\hat{\alpha}_{id}$ is the maximum likelihood estimator of $\alpha_{id}$ defined in Equation \eqref{equ:mle_alpha}, $G(x) = \partial \log \Gamma(x) / \partial x$ is the digamma function, and a score residual $s_{id} = G(\sum_i \hat{\alpha}_{id}) - G(\hat{\alpha}_{id}) + \log v_{id}$. Note that the denominator of $\rho_{id}$ reflects the amount of observed information of $v_{id}$ that  contributes to the estimation of the parameter $\alpha_{id}$ \cite{gueorguieva2008dirichlet}.
    
    Another important diagnostic tool for the Dirichlet regression is {\it overdispersion}, which evaluates the goodness-of-fit of the model. 
    The overdispersion is defined as an increase of the variance of a response due to the lack of homogeneity in a parameter across observations \cite{zelterman1988homogeneity}. Zelterman and Chen derived overdispersion statistics when parameters are allowed to have small amount of random variability in the response across observations \cite{zelterman1988homogeneity}. The test statistics are to detect a lack of fit when the variability is larger than the expected variances.
    For a set of mutually independent response variables $V_i$ having the pdf $\mathcal{D}_i(v_i | \alpha_{i1},\ldots,\alpha_{iD})$, the homogeneity in each parameter $\alpha_{id}$ is satisfied when all $\alpha_{id}$'s are identical across observations. Gueorguieva {\it et al.} (2008) provided test statistics for overdispersion consisting of the shape parameters $\alpha_{id}$ and coefficients $\beta_{dk}$. For the Dirichlet regression model with the $k-th$ covariates $x_k$ for the $i$-th observation, the overdispersion statistic for testing homogeneity of the parameter $\alpha_{id}$ and the regression coefficient $\beta_{dk}$ is defined as 
    \begin{equation}
        \label{eq:delta}
         \delta_{id}^{\beta_{dk}} = \alpha_{id}^2 x_{ik}^2 \eta_{id}^{\alpha_{id}} \quad d=1,2,\dots,D, \quad i=1,2,\dots,n,
    \end{equation}
    with 
    \begin{equation} \label{eq:eta}
     \eta_{id}^{\alpha_{id}} = G'(\sum_i \hat{\alpha}_{id}) - G'(\hat{\alpha}_{id}) + s_{id}^2.
    \end{equation}
   In general, if sample variances of certain observations are larger than what is expected, the estimates for $\delta_{id}^{\beta_{dk}}$ are also getting larger accordingly.
   
    
\section{Results}
\subsection{Log-ratio regression}

In an initial investigation of the variances of the log-ratios, it was found that the rarer components engendered high variances. Hence it was decided to use weighted analyses in the multivariate analyses where the transformed components of the cell types are combined, for example in the computation of total log-ratio variance as in Equation \eqref{eq:TLRV} or the computation of multivariate distances. Here the default weights were the average proportions of the nine immune cell types, so that rarer cell types have less weight than the more abundant ones.
However, notice that the weighting does not affect the log-ratio regressions, where the transformed compositions act as response variables. 

\medskip
\leftline{\it Choice of \alr~transformation}
\smallskip
\noindent In order to select the denominator for the \alr~transformation, Table~\ref{tab:procrustes} presents the Procrustes correlations, in descending order, of the \alr-transformed data using each cell type in turn as denominator. The weights in this table refer to the average proportions in the whole data set, which are used to compute total log-ratio variance in Equation \eqref{eq:TLRV} and the exact log-ratio geometry. In this case, the order of the correlations follows the weight of the denominator component. 
    \begin{table}[htbp!]
        \caption{Correlations from the Procrustes analysis between the geometries of the different \alr~transformations and the exact geometry of all pairwise logratios, using each cell type in turn as denominator.} \label{tab:procrustes}
       \vspace{0.2cm}
      \centering
      \resizebox{.5\textwidth}{!}{
        \begin{tabular}{lrr} \bottomrule
        Cell type       & Weight & Procrustes Correlation \\ \hline
        Macro      & 0.453  & 0.989                 \\
        T.CD8     & 0.142  & 0.920                 \\
        T.CD4     & 0.165  & 0.909                 \\
        Mast      & 0.088  & 0.860                 \\
        B         & 0.072  & 0.851                 \\
        NK        & 0.051  & 0.829                 \\
        Dendr     & 0.017  & 0.689                 \\
        Neutr     & 0.008  & 0.628                 \\
        Eosin     & 0.003  & 0.569                \\ \toprule
        \end{tabular}
        } 
    \caption*{\hspace{3cm}\footnotesize Weight is the average proportion of all log-ratios in the weighted analysis.}
    \end{table}
\noindent    
    The higher the Procrustes correlation, the more accurately the ALR transformation reflects the exact logratio structure. Hence Table~\ref{tab:procrustes} shows that {\it Macrophage} is the reference part of choice, with a Procrustes correlation of 0.989.
    
\begin{figure}[htbp!]
        \centering
        \includegraphics[width=0.99\textwidth]{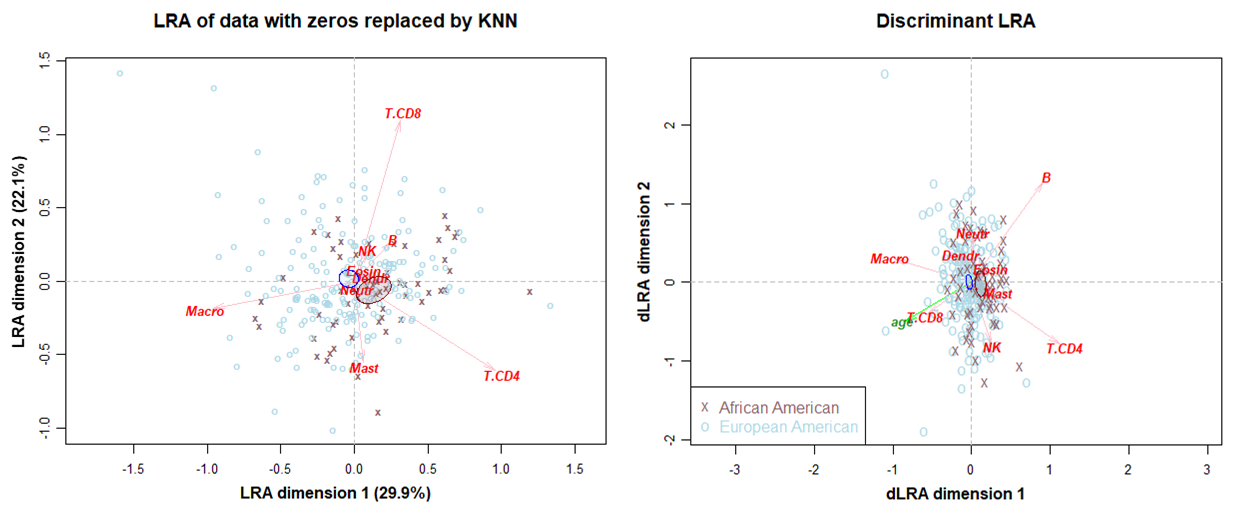}
        \vspace{0.2cm}
        \caption{(Left) Weighted log-ratio analysis of the immune cell compositional dataset, showing the contribution biplot. The African American samples are indicated by a cross, and the two ellipses are 95\% confidence ellipses for the group means, with the African American group on the right. (Right) The discriminant version where the first dimension coincides with the group difference.}
        \label{fig:LRAs}
    \end{figure}

\begin{figure}[htbp!]
        \centering
        \includegraphics[width=0.6\textwidth]{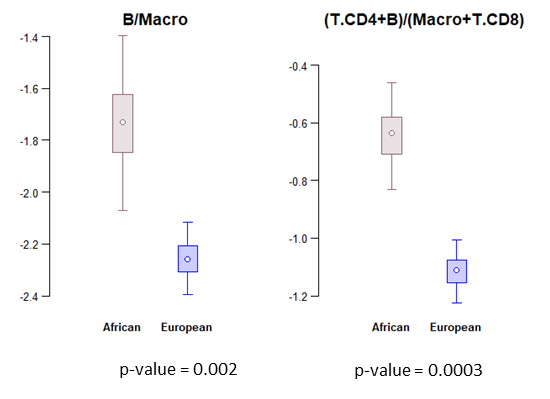}
        \vspace{0.2cm}
        \caption{95\% confidence plots of the log-ratio means of the pairwise log-ratio on the left and the summated log-ratio on the right. The dot indicates the mean, the box indicates the 50\% confidence interval and the whiskers extend to the 95\% confidence interval.}
        \label{fig:ConfPlots}
    \end{figure}

\medskip
\leftline{\it Exploratory multivariate analyses of the immune cell compositions} 
\smallskip
\noindent Before doing the log-ratio regression, it is interesting to understand the multivariate structure of the data set. 
Figure \ref{fig:LRAs} shows the weighted LRA on the left, that is the weighted PCA of the CLR-transformed data, where the two racial groups are coded into the label of the samples. 
This analysis, with $29.9+22.1= 52.0\%$ of the total log-ratio variance explained shows that four immune cell types dominate: Macrophage, T.cells.CD8, T.cells.CD4 and Mast.cells. 
On the right a discriminant version of the structure is shown, where the first (horizontal) axis is specifically constrained to coincide with the difference in the two group means of EA and AA. 
From this latter plot, it can be deduced that log-ratios such as B.cells/Macrophage and T.cells.CD4/T.cells.CD8 are good discriminators of the two groups, or even the amalgamation log-ratios of  (B.cells+T.cells.CD4)/(Macrophage+T.cells.CD8), called SLRs (summated log-ratios, see Greenacre {\it et al.}, 2020). 
This can be demonstrated by making 95\% confidence plots of the group means (Greenacre, 2016) and performing a statistical test of the group differences, shown in Figure \ref{fig:ConfPlots}.
If the confidence intervals do not overlap, this is a reliable indication that the group means are significantly different, but the statistical test should nevertheless be performed to confirm this result.
The p-values are indeed both less than 0.05, with the summated log-ratio having the greater separation and the lower p-value.

\medskip
\leftline{\it Linear regressions using log-ratio transformed responses}
\smallskip
\noindent Each \alr-transformed log-ratio is regressed in turn on race, a dummy variable coding AA, and a continuous variable for age. 
The estimated regression coefficients for these two variables are listed in Table
\ref{tab:multiplicative}, along with their 95\% Bootstrapped confidence intervals (10000 bootstrap replicates).

    \begin{table}[htbp!]
        \caption{Multiplicative coefficients for the the \alr~transformed responses.}
        \vspace{0.2cm}
        \centering
        \resizebox{.8\linewidth}{!}{
        \begin{tabular}{l|cccc}
            \bottomrule														
                        	&	race(AA)	&	boot.CI\textsubscript{race}				&	age	&	boot.CI\textsubscript{age}				\\	\hline
            T.CD8/Macro	&	1.0234	&	(	0.7312	1.3902	)	&	1.0019	&	(	0.9927	1.0114	)	\\	
            T.CD4/Macro	&	1.6476	&	(	1.2230	2.2024	)	&	1.0004	&	(	0.9889	1.0138	)	\\	
            B/Macro	&	1.6617	&	(	1.1485	2.3378	)	&	0.9922	&	(	0.9811	1.0032	)	\\	
            NK/Macro	&	1.3929	&	(	1.0139	1.8682	)	&	1.0032	&	(	0.9938	1.0121	)	\\	
            Dendr/Macro	&	1.0676	&	(	0.6843	1.5620	)	&	0.9976	&	(	0.9872	1.0086	)	\\	
            Mast/Macro	&	1.3601	&	(	0.9220	1.8937	)	&	0.9996	&	(	0.9899	1.0093	)	\\	
            Neutro/Macro	&	1.1689	&	(	0.8237	1.6187	)	&	0.9944	&	(	0.9834	1.0055	)	\\	
            Eosin/Macro	&	1.4902	&	(	1.0052	2.1052	)	&	0.9969	&	(	0.9874	1.0067	)	\\	\toprule
            \end{tabular}
        }
         \caption*{\hspace{1.5cm}\footnotesize Note that boot.CI\textsubscript{race} and boot.CI\textsubscript{age} denote the 95\% bootstrap confidence intervals for race and age.}
        \label{tab:multiplicative}
    \end{table}
    The coefficients present how much the \alr~response is influenced multiplicatively by the explanatory variable. 
    For example, for AA the ratio B.cells/Macrophage is estimated as 66\% higher than the same ratio for EA. The fact that the confidence interval does not include 1 means that this is a significant result, which has already been seen noted in Figure \ref{fig:ConfPlots}. Two other \alr~ratios, NK.cells/Macrophage and Eosinophils/Macrophage, have significant coefficients, showing 39\% and 49\% increase in AA, respectively. The coefficients of age for all responses are all close to 1, and their confidence intervals all include 1, indicating that there is no significant effect of age on the responses.

    \begin{figure}[htbp!]
        \centering
        \includegraphics[width=0.9\textwidth]{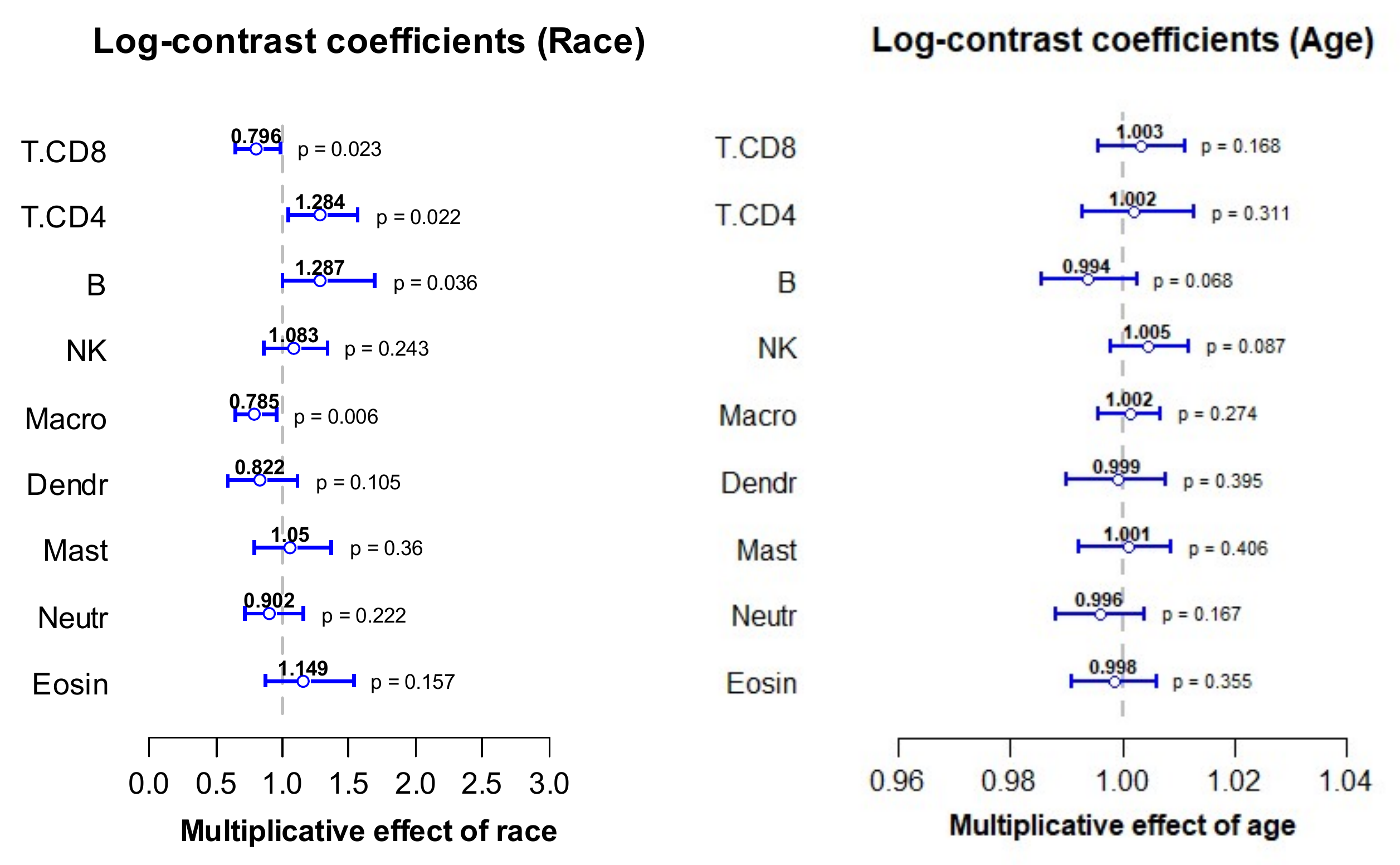}
        \vspace{0.2cm}
        \caption{Estimates of log-contrast coefficients (exponentiated) for each immune cell type, along with 95\% bootstrap confidence intervals and p-values. Race is coded as a dummy variable for African American.}
        \label{fig:LogContrasts}
    \end{figure}
    The estimates for the \alr~responses can be converted into log-contrast coefficients estimated for individual compositional components.
    Figure \ref{fig:LogContrasts} shows these coefficients, also expressed as multiplicative effects, their bootstrap confidence intervals and p-values, for both race and age.
    Once again, for age, all the coefficients are close to 1 and none significantly different from 1, indicating no effect.
    For race, four components are significant, two in a positive direction (i.e. multiplicative effects greater than 1, T.cells.CD4 and B.cells) and two in a negative direction (T.cells.CD8 and Macrophage), coinciding exactly with the results on the right of Figures \ref{fig:LRAs} and \ref{fig:ConfPlots}.
    These individual coefficients for race can be interpreted as follows: given any particular composition $\bf v$ of the nine cellular types for EA, the corresponding composition for AA is the set of estimated coefficients multiplied elementwise with $\bf v$, hence pushing some values upwards, others downwards.
    Since race and age have been included in the regressions, the effects of age would also have to be applied if the hypothetical AA were different in age, but this will have only small non-significant effects. 
    The beauty of the log-contrast coefficients is that exactly the same result would be obtained for any \alr~transformation, any \ilr~transformation, or the \clr~transformation. The only difference between the transformations is then the estimated coefficients in a table such as Table 2, where it is preferred to have simple and easily interpretable responses.
 
 \begin{table}[h]
    \caption{MANOVA table for model using \alr~transformed immune cells as multivariate responses.} \label{tab:manova}
    \centering
    \resizebox{.5\textwidth}{!}{
    \begin{tabular}{l | rrrr }
        \bottomrule											
        	&	Pillai	&	(df1, df2)	&	approx.F	&	\multicolumn{1}{l}{$p$-value}		\\	\hline
        race(AA)	&	0.0866	&	(8, 244)	&	2.8928	&	0.0043  \\	
        age	&	0.021	&	(8, 244)	&	0.6528	&	0.7327  \\	\toprule
    \end{tabular}											
    }
    \caption*{\footnotesize Pillai is Pillai's trace, approx.F is the approximation of Pillai's trace on a F-statistic and (df1, df2) are the degrees of freedom of the approximated F-statistics.}
\end{table}

 \begin{figure}[htbp!]
    \centering
    \includegraphics[width=0.4\textwidth]{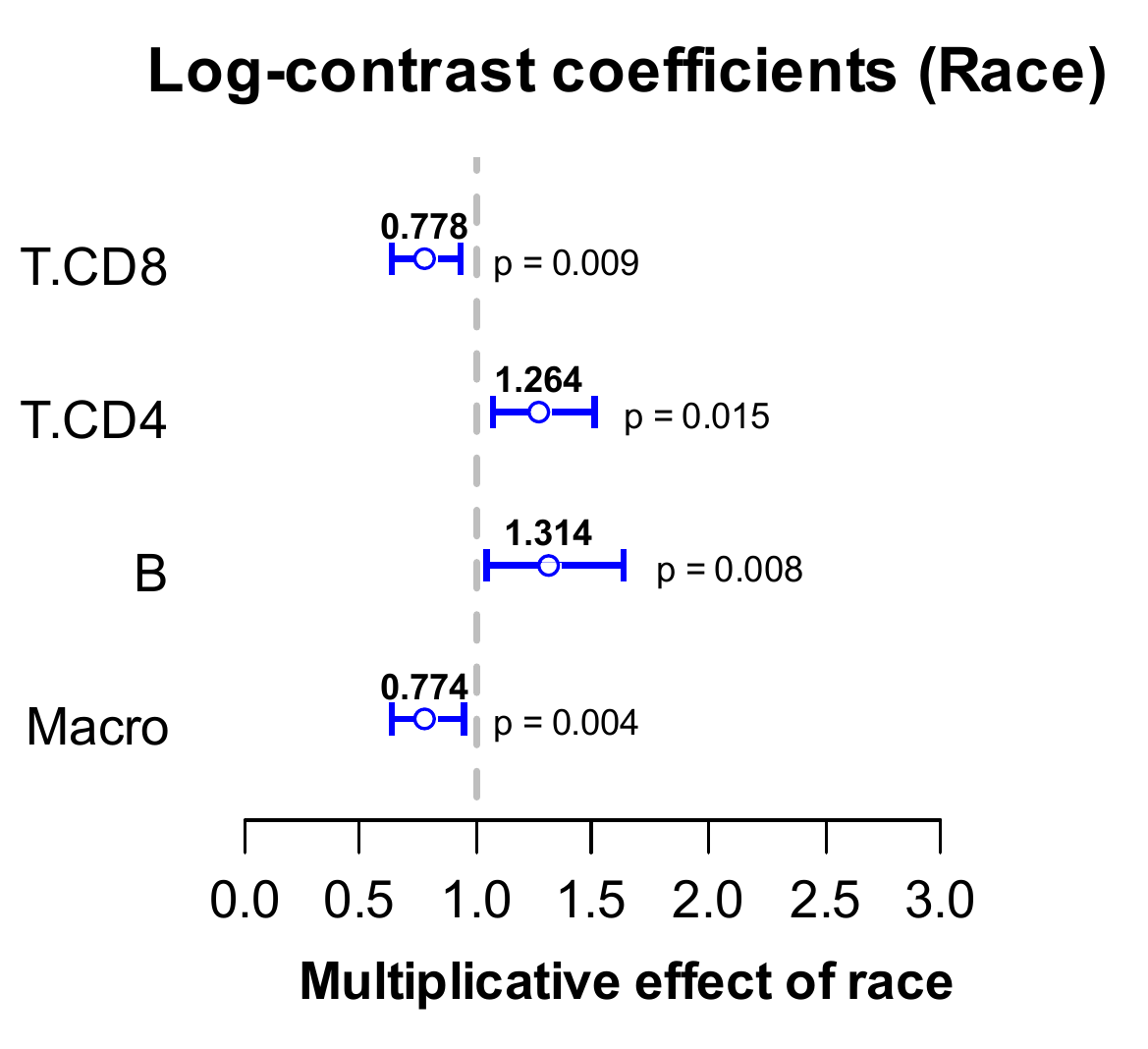}
    \vspace{0.2cm}
    \caption{Estimates of log-contrast coefficients (exponentiated) for the four-part subcompositional repsonse modelled on the discrete predictor race, with a dummy variable for the category African American, along with 95\% bootstrap confidence intervals and p-values.}
    \label{fig:Subcomposition}
\end{figure}
  An alternative way to conclude that age is not a significant predictor of the compositional response is to conduct the MANOVA test for the models. Table~\ref{tab:manova} shows the results for \alr~model with Pillai's trace and approximated F-statistics with its degree of freedoms. Pillai's trace value ranges from 0 to 1, which indicates that the explanatory variable has a significant effect on the multivariate response as being closer to 1 (Pillai, 1955). The result in Table~\ref{tab:manova} shows that race is significant on immune cells with Pillai's trace of 0.0866 ($p = 0.0043$). On the other hand, age is not significant.

    In regression analysis, to arrive at a final model, the nonsignificant predictor variable of age should be omitted. 
    Furthermore, in this case where the composition is the multivariate response, the nonsignificant components of the response can also be eliminated, arriving at a parsimonious description of the relationship.  
    The logratio regressions were thus repeated including only four immune cell types: T.cells.CD8, T.cells.CD4, B.cells and Macrophage, as a subcomposition. 
    The results for the log-contrast coefficients are given in Figure \ref{fig:Subcomposition}, showing estimates, 95\% confidence intervals and p-values.
    The results are more significant on each of the cell types, as might be expected in this reduced model.  
    Furthermore, the similarity of the multiplicatively increasing coefficients for T.cells.CD4 and B.cells and those of the multiplicatively decreasing coefficients for T.cells.CD8 and Macrophage, gives further credence for using the simple ratio (T.cells.CD4+B.cells)/(T.cells.CD8+Macrophage), as already shown in Figure 3, deduced from the right hand biplot of Figure 2.

\subsection{Dirichlet regression}
    Since the age covariate is insignificant, we only focus on a Dirichlet regression model with race.
        
\medskip
\leftline{\it Estimates of regression coefficients}
\smallskip
    \noindent Table~\ref{tab:dirichlet} provides estimates of the $\beta$ coefficients, standard errors (s.e.) and p-values for testing $H_0: \beta=0$ vs. $H_1:\beta \neq 0$. The coefficients for T.cells.CD4 and Macrophage are significant, which means that AA has $\exp(0.2127)=1.2371$ times or $\exp(-0.2290)=0.7953$ times more T.cells.CD4 or Macrophage than EA. The signs of the coefficients are generally agreeing with those in Figure \ref{fig:LogContrasts}, where positive and negative coefficients correspond to multiplicative effects above and below 1, respectively.
    \begin{table}[htbp!]
    \centering
    \caption{Dirichlet regression outputs with race as an independent variable.} 
    \label{tab:dirichlet}
    \resizebox{.6\textwidth}{!}{
       \begin{tabular}{l|rrrr}										
        \bottomrule										
        Cell type	&	\multicolumn{1}{c}{estimate}	&	\multicolumn{1}{c}{$\exp(\text{estimate})$}	&	\multicolumn{1}{c}{s.e.}	&	\multicolumn{1}{c}{$p$-value}	\\	\hline
        T.CD8	&	-0.1946	&	0.8232	&	0.1098	&	0.0762	\\	
        T.CD4	&	0.2127	&	1.2371	&	0.1032	&	0.0392	\\	
        B	&	0.2051	&	1.2277	&	0.1175	&	0.0809	\\	
        NK	&	0.0440	&	1.0450	&	0.1217	&	0.7180	\\	
        Macro	&	-0.2290	&	0.7953	&	0.0938	&	0.0147	\\	
        Dendr	&	-0.0924	&	0.9117	&	0.1345	&	0.4920	\\	
        Mast	&	0.0371	&	1.0378	&	0.1158	&	0.7490	\\	
        Neut	&	-0.0377	&	0.9630	&	0.1353	&	0.7810	\\	
        Eosin	&	0.0660	&	1.0682	&	0.1370	&	0.6300	\\	\toprule
        \end{tabular}
    }
    \end{table}

    

    \begin{figure}[htbp!]
    \centering
        \includegraphics[width = 0.8\textwidth]{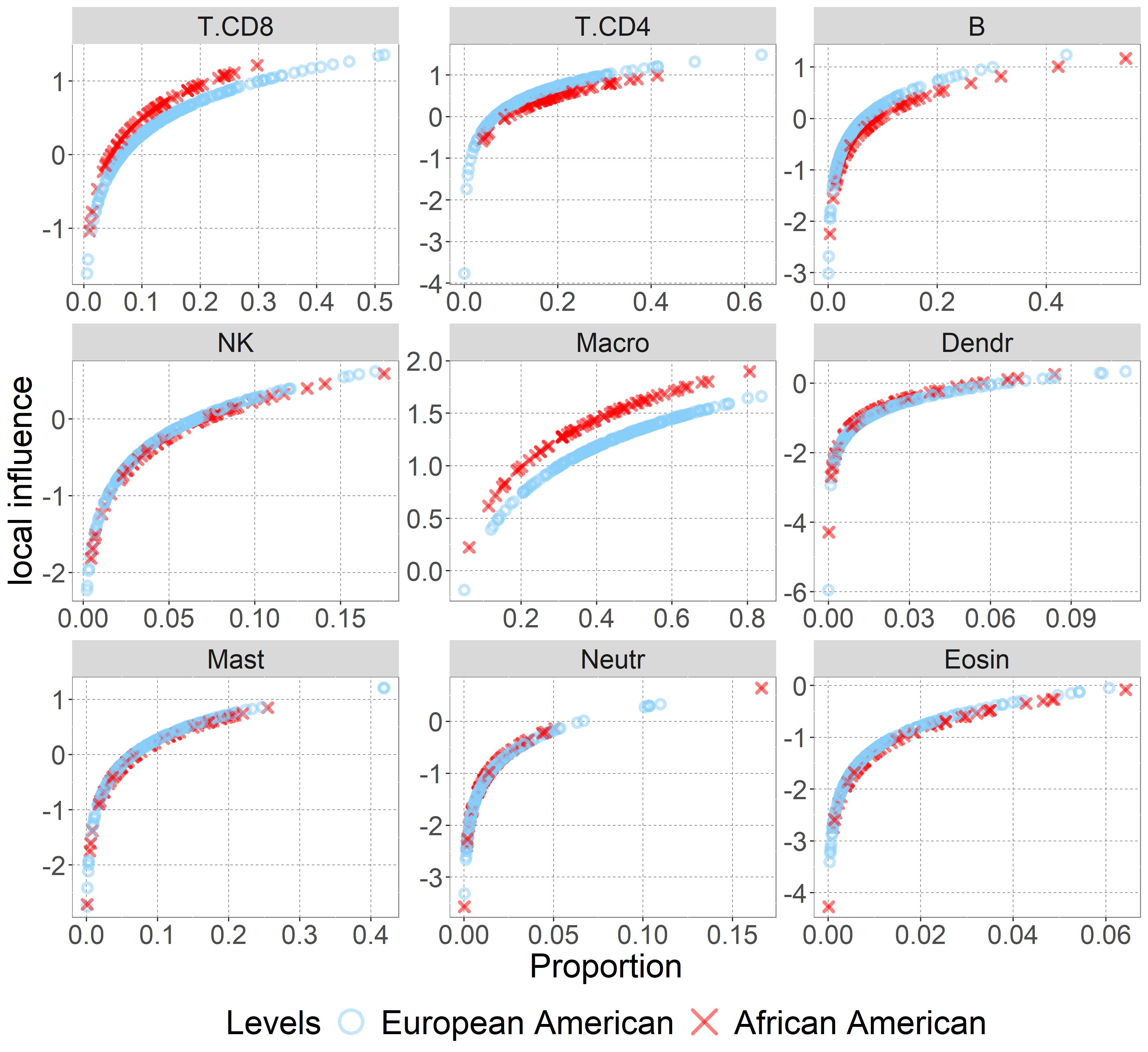}
        \caption{Componentwise plots of the local influence measures against compositional values based on the Dirichlet model with the race variable.}
        \label{fig:dirich_cooks_model(race)}
    \end{figure}
    Figure~\ref{fig:dirich_cooks_model(race)} is a componentwise plot of the local influence measures against compositional values, which were generated based on the fitted Dirichlet models with race as an independent variable. We have already verified that race has no significant relevance with the immune cells in Table~\ref{tab:dirichlet}. Overall, the local influence measures tend to increase rapidly for values near zero and then increase more gradually as values increase. In spite of varying curvatures among cell types, which is smallest for Macrophage, it is common that as values are getting close to zero, the impact of individual observations on the estimation increases significantly. In Figure~\ref{fig:dirich_cooks_model(race)}, we can also find two curves for T.cells.CD8, T.cells.CD4, B.cells, NK.cells and Macrophage corresponding to racial groups, which diverges. Specifically, AA has larger effects on estimates for Macrophage and T.cells.CD8 compared to EA, whereas opposite directionalities are observed for T.cells.CD4, B.cells and NK.cells. On the other hand, two curves are not visually separable for Dendritic.cells, Mast.cells, Neutrophils and Eosinophils.

\begin{figure}[htbp!]
        \centering
        \includegraphics[width=.5\textwidth]{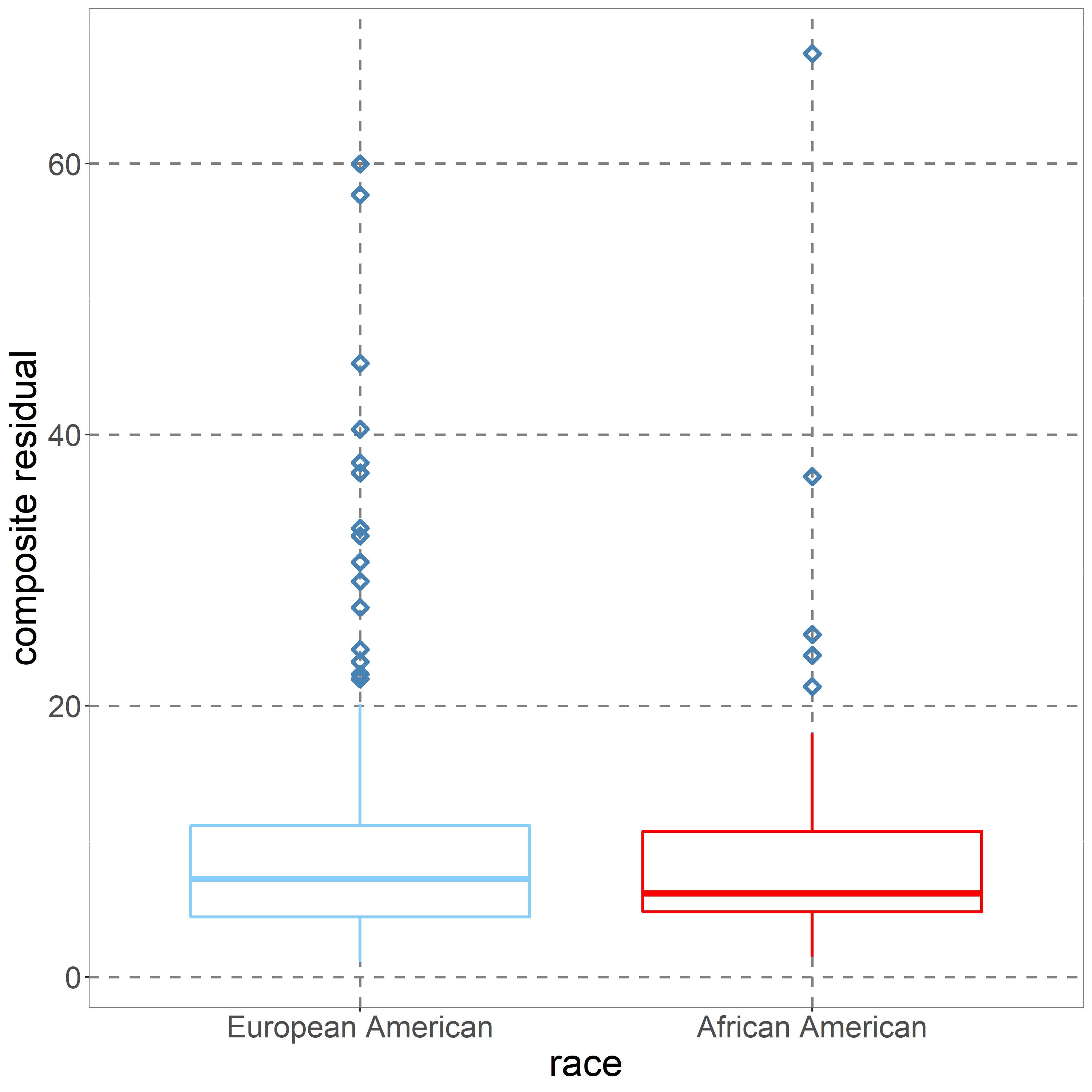}
        \caption{Composite residual plot for Dirichlet regression model with race.}
        \label{fig:dirich_cmpresid}
    \end{figure}

   Figure~\ref{fig:dirich_cmpresid} illustrates the composite residual plots of the Dirichlet model, which shows that there are some observations with large composite residuals over 40 in both racial groups, but majority of the composite residuals are spread below 20.
        
\begin{figure}[htbp!]
        \centering
        \includegraphics[width = .9\textwidth]{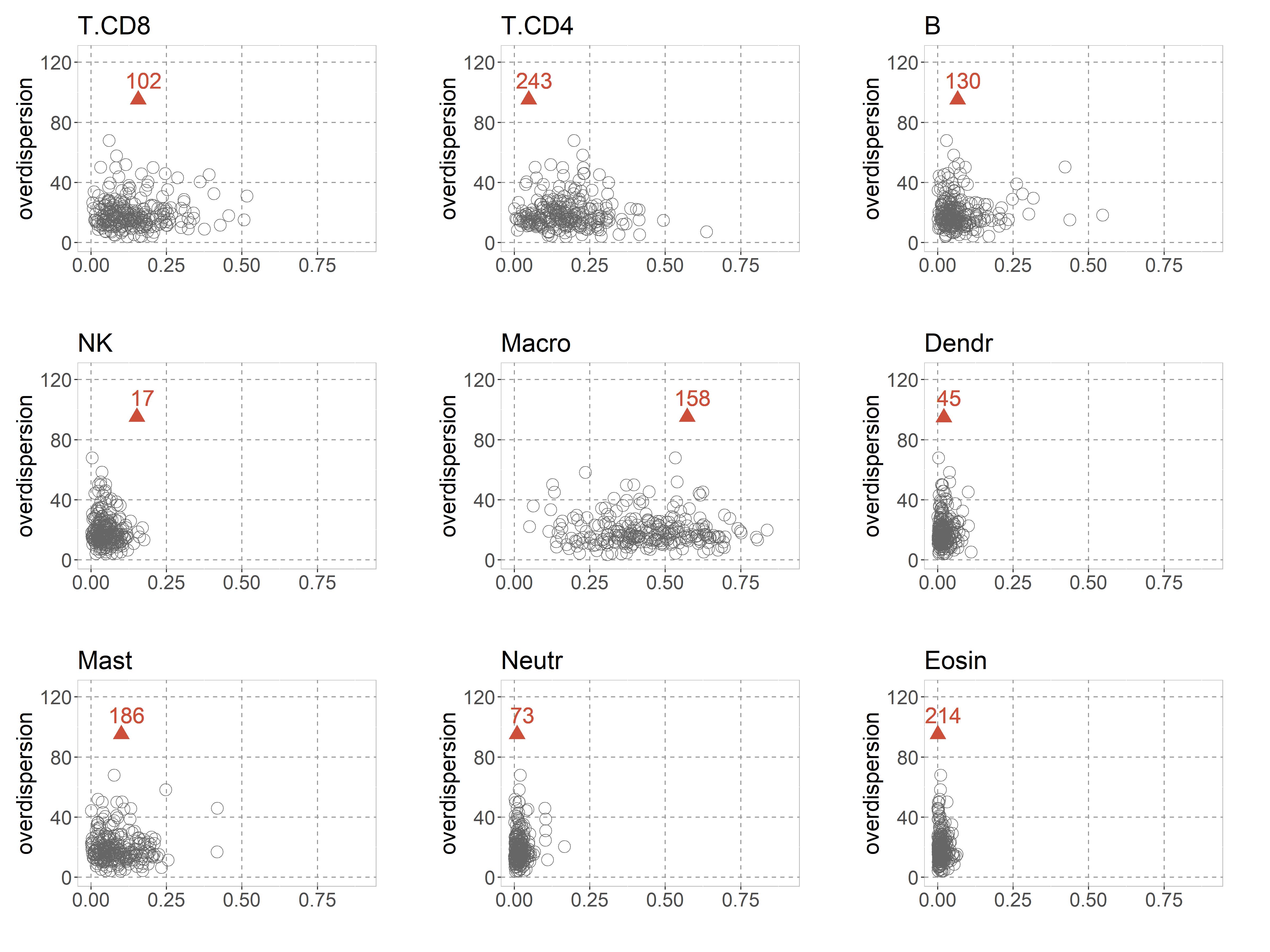}
        \caption{Componentwise plots of the overdispersion statistic against compositional values based on the Dirichlet model fitted with the race variable. The red marked points indicate the observation with the largest overdispersion statistic value in each cell type.}
        \label{fig:dirich_overdisp_model(race)}
    \end{figure}

    Figure~\ref{fig:dirich_overdisp_model(race)} illustrates the componentwise plots of the overdispersion statistics of individual observations against compositional values, based on the Dirichlet models fitted with race. In these plots, the red marked points indicate the observation with the largest overdispersion statistic value in each cell type. 
    Nonetheless, no significant overdispersion issue is detected in general.
    
\section{Discussion}
\label{sec5}

With improved understanding of interaction between the immune system and various diseases such as cancer, the immunology field studying human immune system has gained significant attention. Investigation of immune cellular composition and its association with diseases constitutes the core of the immunologic studies. However, in spite of their importance, optimal statistical strategies for this type of data still remain to be studied. In this paper, we reviewed statistical methods for compositional data analysis and applied the methods to colorectal cancer immune cellular fractions data.

As illustrated throughout the manuscript, it is critical to consider unique aspects of compositional data to implement efficient data analysis of immune cellular composition data and guarantee meaningful scientific insight. Ignoring this can result in misleading conclusions based on inappropriately visualization and/or suboptimal selection of key variables ignoring inter-relationships among the elements in compositional data. As solutions for these issues, we especially investigated the log-ratio and Dirichlet regression models. Each approach has its own strengths. One of the key strengths of the log-ratio approaches is the fact that existing and established statistical methods can be employed. This allows utilization of a wide range of existing statistical models. 
The log-ratio approach involves choosing one of the available log-ratio transformations, which in the present application serve as multivariate responses in a regression model. Fortunately, for this purpose the final results in the form of log-contrast coefficients are invariant with respect to this choice, so we have used the simplest option, \alr. This choice has the favourable property that the individual regressions can be more easily interpreted.
In contrast, the Dirichlet model handles compositional data more directly, without transformation and with a simpler interpretation, but the analysis is no longer subcompositionally coherent. 

In the analysis of colorectal cancer immune cellular fractions data, we mainly focused on studying associations of immune cellular fractions with race, since age was found to not affect the responses significantly in both analyses. 
The log-ratio regression found that four cellular types were significantly associated with the two racial groups, whereas the Dirichlet regression found only two of those four types to be significant.
From the log-ratio regression we can conclude that T.cells.CD4, B.cells, T.cells.CD8 and Macrophage can potentially be considered as key markers for racial difference, and that the ratio of the sum of first two versus the sum of the last two can be used as a single summary of the distinction between the two groups. 

We hope that this paper provides a gentle but thorough guideline for the statistical analysis of compositional data, especially those generated in immunology.
\section*{Data availability}
The data we used in this paper is available as Table S2 of The Immune Landscape of Cancer paper (Thorsson {\it et al.}, 2018, Immunity, 48: 812-830; \url{https://www.sciencedirect.com/science/article/pii/S1074761318301213#app2}). Corresponding clinical information is available in the cBioPortal website (\url{http://www.cbioportal.org/}).

\section*{Acknowledgement}
This work was supported by the National Institutes of Health (grant numbers R01-GM122078, R21-CA209848, U01-DA045300, U54-AG075931) awarded to Dongjun Chung, and the National Research Foundation of Korea (NRF-2019R1F1A1061691) awarded to Young Min Kim. The funders had no role in study design, data collection and analysis, decision to publish, or preparation of the manuscript.

\hspace{-0.4cm}\textit{Conflict of Interest:} None declared.
\vspace*{-12pt}

\bibliographystyle{plain}
\bibliography{ref}

\end{document}